\documentstyle[12pt]{article}
\begin{document}
\title{A New Theory of Geomagnetism}
\author{B.G. Sidharth$^*$\\
Centre for Applicable Mathematics \& Computer Sciences\\
B.M. Birla Science Centre, Adarsh Nagar, Hyderabad - 500 063 (India)}
\date{}
\maketitle
\footnotetext{$^*$Email:birlasc@hd1.vsnl.net.in; birlard@ap.nic.in}
\begin{abstract}
It is pointed out, that in the light of recent results on the semionic
or anomalous behaviour of electrons below the Fermi temperature, the
solid core of the earth which has been ignored so far, would contribute
significantly to Geomagnetism and help explain the puzzling magnetic
reversals.
\end{abstract}
\section{Introduction}
In 1905 Albert Einstein described Geomagnetism as one of the five unsolved
problems of physics\cite{r1}. After nearly a century, inspite of a tremendous
amount of work which has culminated in the dynamo model of Geomagnetism,
\cite{r2}, it cannot be said with confidence that the problem has been
solved. From time to time, simulations are developed which improve upon
earlier models\cite{r3,r4,r5}, but several unexplained features persist.
These include the problems of Geomagnetic reversals\cite{r6,r7,r8,r9}. In
particular it may be mentioned (cf.ref.\cite{r9} and \cite{r1})that Muller and
Morris have attributed the reversals to asteroid impacts, which in turn
have been related to mass extinctions.\\
We will point out in what follows that in the light of recent work on
semionic and anomalous behaviour of Fermions under special conditions, the
solid core of the earth which has hitherto not been considered could
contribute significantly to Geomagnetism, and could even facilitate an
explanation for the magnetic reversals.
\section{Magnetism of the Solid Core}
It is well known\cite{r10,r11} that the earth has a solid core with a radius
of about 1200 kilometers, composed mostly of Iron $(90\%)$ and Nickel $(10\%)$,
at a temperature of about 6000 degrees centigrade and with a relative
density around 10. This in turn is surrounded by the liquid core which
it is believed gives rise to the dynamo model of Geomagnetism.\\
Given the above data, using the atomic weight of iron, we can easily
calculate that the number of atoms in the solid core, $N$ is given by,
\begin{equation}
N \approx 10^{48}\label{e1}
\end{equation}
where the symbol $\approx$ denotes, "of the order of".\\
We next calculate the Fermi temperature of the conduction electrons in the
solid core. This is given by\cite{r12},
\begin{equation}
kT_F = \left\{6\pi^2\left(\frac{N}{V}\right)\right\}^{2/3} \frac{\hbar^2}{2m}\label{e2}
\end{equation}
where $V$ is the volume of the solid core, $k$ is the Boltzmann constant
$\hbar$ is the reduced Planck constant and $m$ the electron mass.\\
It follows from  (\ref{e2}) that
\begin{equation}
T_F \approx 10^{5^o}C\label{e3}
\end{equation}
Thus one can see that the temperature of the solid core is below the
Fermi temperature of the conduction electrons.\\
In recent years it has been realized that under special conditions like low
dimensionality or temperatures, conduction electrons do not strictly obey
Fermi-Dirac statistics, but rather they are semionic, that is
they obey statistics between the Fermi-Dirac and Bose-Einstein statistics
\cite{r13,r14}. In particular, this is true below the Fermi temperature
\cite{r15}. The implications are interesting:\\
Given Fermi Dirac statistics, at temperatures below the Fermi temperature
we would have (cf.ref.\cite{r12}) for the magnetisation $M$ per unit volume
the formula
\begin{equation}
M = \frac{\mu (2\bar N_+ - N)}{V}\label{e4}
\end{equation}
where $\mu$ is the electron magnetic moment and $\bar N_+$ is the
average number of electrons with spin up, say, where $\bar N_+ \approx
\frac{N}{2}$, so that $M$ in (\ref{e4}) is very small.\\
However if the behaviour is not Fermionic, but rather Bosonic,
then
$$\bar N_+ \approx N$$
In our case,
$$\frac{N}{2} < \bar N_+ < N$$
With this input (\ref{e4}) becomes,
\begin{equation}
M \leq \frac{\mu N}{V}\label{e5}
\end{equation}
From (\ref{e5}) we can easily deduce that the terrestrial magnetic field
$H$ is given by,
$$H \leq \frac{MV}{r^3} \approx 1 G$$
The above order of magnetic calculation thus gives the correct order of the
terrestrial magnetism. Moreover, this would have the added advantage that
it could explain geomagnetic reversals: The semionic behaviour of the electrons
in the solid core is sensitive to external magnetic influences and could
thus flip or reverse polarity. On the other hand, the explanation in the
case of the convective dynamo model would be contrived in comparison.

\end{document}